\documentclass[prd,twocolumn,floats,floatfix,showpacs,nofootinbib]{revtex4}
\usepackage{graphicx}
\usepackage{dcolumn}
\usepackage{bm}
\usepackage{graphics}
\usepackage{slashed}
\usepackage{amssymb}
\usepackage{natbib}
\usepackage{amsmath}
\usepackage{amsthm}
\usepackage{url}
\usepackage{color}
\newcommand{\bea}{\begin{eqnarray}}
\newcommand{\ena}{\end{eqnarray}}
\newcommand{\bean}{\begin{eqnarray*}}
\newcommand{\enan}{\end{eqnarray*}}

\newcommand{\fracc}[2]{\frac{\textstyle{#1}}{\textstyle{#2}}}
\newtheorem*{theorem}{Lemma}
\newtheorem*{hypothesis}{Fundamental Hypothesis}
\begin{document}

\title{Geometric scalar theory of gravity}

\author{M. Novello$^1$\footnote{M. Novello is Cesare Lattes ICRANet
Professor}}
%\email{novello@cbpf.br}
\author{E. Bittencourt$^1$} %\email{eduhsb@cbpf.br}
\author{U. Moschella$^2$} %\email{Ugo.Moschella@uninsubria.it}
\author{E. Goulart$^1$} %\email{egoulart@cbpf.br}
\author{J. M. Salim$^1$} %\email{jsalim@cbpf.br}
\author{J. Toniato$^1$} %\email{toniato@cbpf.br}

\affiliation{$^1$Instituto de Cosmologia Relatividade Astrofisica ICRA -
CBPF\\
Rua Dr. Xavier Sigaud 150 - 22290-180 Rio de Janeiro - Brazil\\}

\affiliation{$^2$Universit\`a degli Studi dell'Insubria - Dipartamento di Fisica e Matematica\\
Via Valleggio 11 - 22100 Como - Italy and\\
INFN, Sez di Milano, Via Celoria 16,
20146, Milano - Italy}


\date{\today}

\begin{abstract}
We present a geometric scalar theory of gravity. Our proposal will
be described using the ``background field method" introduced by
Gupta, Feynman, Deser and others as a field theory formulation of general
relativity. We analyze previous criticisms against scalar gravity
and show how the present proposal avoids these difficulties. This
concerns not only the theoretical complaints but also those related
to observations. In particular, we show that the widespread belief
of the conjecture that the source of scalar gravity must be the
trace of the energy-momentum tensor---which is one of the main
difficulties to couple gravity with electromagnetic phenomenon in
previous models---does not apply to our geometric scalar theory.
Some consequences of the new scalar theory are explored.
\end{abstract}

 \maketitle

\section{Introduction}

In recent years a few papers have appeared reviewing
the scalar theory of gravity and pointing out the reasons
for which that theory was dismissed
(see \cite{gibbons}, \cite{giulini} and \cite{will} and
references therein for a fairly complete review on scalar
theories of gravity). Criticisms range from theoretical
to observational; they are based on the hypothesis made since the old days of the Einstein-Grosmann
proposal \cite{coll-eins} that scalar gravity deals with
essentially three assumptions:
\begin{itemize}
\item{The theory is described in a conformally flat geometry and the background
Minkowski metric is observable;}
\item{The source of the gravitational field is the trace of the
energy-momentum tensor;}
\item{The scalar field is the (special) relativistic generalization of the Newtonian potential.}
\end{itemize}

In this paper we present a new possibility for describing
the gravitational interaction in terms of a scalar field $\Phi$,
i.e. a geometric scalar gravity (GSG). The above three assumptions
do not hold in our theory. In particular, we shall see that
the general belief that the only possible source of scalar
gravity is the trace of the energy-momentum tensor is not valid here.

Let us clarify from the very beginning that this is not a special
relativistic scalar gravity. The adjective geometric in the title
pinpoints its similarity with general relativity: we are
constructing a metric theory of gravity. In other words, we follow
the main idea of general relativity and assume as an \textit{a
priori} that gravity is described by a Riemannian geometry. In
general relativity the ten components of the metric tensor are the
basic variables of the theory (up to coordinate transformations).
Here the metric tensor is determined by the derivatives of a
fundamental independent physical quantity represented by the scalar
field $ \Phi.$

This means that although we make use of a scalar field to represent
all gravitational processes we do not follow the previous examples
of scalar gravity, as for instance the Einstein-Grossman ``Entwurf
Theory".  Before presenting the reasons that motivated us to
undertake this proposal, let us summarize the main properties of the
GSG:

\begin{itemize}
\item{The gravitational interaction is described by a
scalar field $ \Phi$;}
\item{The field $\Phi$ satisfies a nonlinear dynamics;}
\item{The theory satisfies the principle of general covariance.
In other words, this is not a theory restricted to the realm of special relativity;}
\item{All kind of matter and energy interact with $ \Phi$
only through the pseudo-Riemannian metric
\begin{equation}
q_{\mu\nu} = a \, \eta_{\mu\nu} + b \,
\partial_{\mu}\Phi \,\partial_{\nu} \Phi;
\label{112} \end{equation}}
\item{Test particles follow geodesics relative to the gravitational metric $ q_{\mu\nu};$}
\item{ $\Phi$ is related in a nontrivial way with the Newtonian potential $\Phi_N$;}
\item{Electromagnetic waves propagate along null geodesics relative to the metric $ q^{\mu\nu}.$}
\end{itemize}

The parameters $ a $ and $ b$ are functionals of the scalar field $
\Phi$ (which will be specified by fixing the Lagrangian of the
scalar field). The auxiliary (Minkowski) metric $ \eta^{\mu\nu}$ is
unobservable because the gravitational field couples to matter only
through $ q^{\mu\nu}$. Here we follow the main steps of general
relativity where a unique geometrical entity interacts with all
forms of matter and energy and the geometry underlying all events is
controlled by the gravitational phenomena.

From this postulate it follows immediately that the space-time geometry
is an evolutionary process identified to the dynamics of the gravitational field.
This beautiful hypothesis of general relativity is contained in each observation
as a specific example of a geometry solving that dynamics.

The second important postulate of general relativity states that
the metric couples universally and minimally to all fields of the
standard model by replacing everywhere the Minkowski metric $
\eta_{\mu\nu}$ by $ g_{\mu\nu}.$

We will accept also this postulate, but we investigate
a special form for the Riemannian metric that
represents the gravitational field. We shall describe the origin of
this departure from general relativity in a later section after reviewing in the next section the case of general
relativity.

A final remark concerns our method to identify the
right form of the dynamics of the scalar field. The action
of a given theory is generally constructed using certain a
priori principles and rules. For instance one might want
to impose general covariance, symmetry principles such
as the gauge principle and a limitation on the order of
derivatives. Although these principles are preserved in
our theory they are of course not enough. Here we adopt
observation-oriented procedure and try to determine the
form of the dynamics of our GSG as a power expansion
in the field $\Phi$ whose coefficients are fixed from observations.
This means that we try to proceed step by step,
adding new terms to the Lagrangian when observation
imposes this, and we continue such procedure until the
theory achieves its final form. In the
following we present an example of how this strategy can
be used by the analysis of the motion of test particles
(like planetary orbits) in the field generated by a massive
body like the sun in the quasi-linear regime.

\section{The description of general relativity as a field theory}

Although general relativity is usually presented in the framework of
Riemannian geometry, it is possible to fully describe the exact
Einstein\rq s theory of gravity in terms of a spin-2 field
propagating in an arbitrary background space-time. The case of
Minkowskian background was investigated by many authors (see details
in \cite{gupta} and \cite{feynman}) and for a generic background
including Minkowski it was described in \cite{deser} and \cite{gpp}.

The main idea can be summarized as follows. Consider a flat
Minkowski background (just to simplify our exposition) endowed with
a metric $\eta_{\mu\nu}.$ In a Lorentzian coordinate system the
metric of the background takes the standard constant expression. We
may allow for general coordinates; the curvature tensor however
vanishes:
$$R^{\alpha}{}_{\beta\mu\nu}(\eta_{\varepsilon\sigma}) = 0. $$
In a galilean coordinate system the metric of the background can
assume standard constant expression. From now on $\eta_{\mu\nu}$ are
the component (not necessarily constant) of the Minkowski metric in
general coordinates. Then one introduces a symmetric second order
tensor $ h_{\mu\nu}$ and writes

\begin{equation}
g^{\mu\nu} \equiv \eta^{\mu\nu}+ h^{\mu\nu}. \label{9julho12}
\end{equation}
This binomial form is an exact expression for the  metric $
g^{\mu\nu}.$ Note however that its inverse, the covariant tensor
$g_{\mu\nu}$ is not in general a binomial form but instead is an
infinite series:
$$ g_{\mu\nu} = \eta_{\mu\nu} -  h_{\mu\nu} + h_{\mu\alpha} \,
h^{\alpha}{}_{\nu} + ... $$

There are two main postulates founding general relativity:
\begin{itemize}
\item{The background Minkowski metric
is not observable. Matter and energy
interact gravitationally only through the combination $
\eta^{\mu\nu}+ h^{\mu\nu}$ and its derivatives. Any test body in
a gravitational field moves along a geodesic relative to the metric $g_{\mu\nu}$;}
\item{The dynamics of gravity is described by an equation relating
the contracted curvature tensor $ R_{\mu\nu} $ to the stress-energy
tensor of matter.}
\end{itemize}

The scalar theory of gravity that we present here deals with a
modification of the flat Minkowski metric similar to the general
relativistic one. However, there is a very important distinction
concerning the origin of the curvature of the space-time and its dynamics. In our
model the (binomial) form of the metric $ q_{\mu\nu} $
comes from a principle that we shall now explain.

\section{The birth of geometry in  scalar gravity}

In this section we will show that a metric $q_{\mu\nu}$
naturally appears in certain nonlinear scalar field theories. Let
us start by considering the following nonlinear Lagrangian in  flat Minkowski spacetime:
\begin{equation}
\label{lagr}
L = V(\Phi) \,w,
\end{equation}
where $w\equiv\eta^{\mu\nu} \partial_{\mu}\Phi \, \partial_{\nu}
\Phi$. For $V = 1/2$ this is just the standard free massless
Klein-Gordon scalar field. In the general case the usual kinetic
term is rescaled by a field dependent amplitude (potential) $V
(\Phi)$. Here we are using $\eta^{\mu\nu}$ but we could have used an
arbitrary coordinate system as well, since the theory is generally
covariant and there is no privileged reference frame. The field
equation is
\begin{equation}
\frac{1}{\sqrt{-\eta}}\partial_{\mu} \, \left(\sqrt{-\eta}  \, \eta^{\mu\nu}\,
\partial_{\nu} \Phi\right) + \frac{1}{2} \, \frac{V'}{V} \, w =0,
\label{23julho3}
\end{equation}
where $ V' \equiv dV/d\Phi $ and  $ \eta $ is the determinant of $
\eta_{\mu\nu}.$

Now comes a remarkable result:
the above field equation (\ref{23julho3}) can be seen as that
of a massless Klein-Gordon field propagating in a curved
space-time whose geometry is governed by $\Phi$ itself.
In other words, the same dynamics can be written
either in a Minkowski background or
in another geometry constructed in
terms of the scalar field. Following the steps established in
\cite{novelloetal}, we introduce the contravariant metric tensor
$ q^{\mu\nu}$ by the binomial formula
\begin{equation}
q^{\mu\nu} = \alpha \, \eta^{\mu\nu} + \frac{\beta}{w} \,
\partial^{\mu}\Phi \,
\partial^{\nu} \Phi,
\label{9junho1}
\end{equation}
where $ \partial^{\mu} \Phi \equiv \eta^{\mu\nu}\,\partial_{\nu} \Phi$
and parameters $\alpha $ and $\beta$ are dimensionless
functions\footnote{Note that the quantity $w$ can be written in
terms of its corresponding $$\Omega \equiv
 \partial_{\mu}\Phi \, \partial_{\nu} \Phi\, q^{\mu\nu}, $$ once one
fixes the theory by specifying the Lagrangian. Indeed, we have
$$ \Omega = (\alpha + \beta) \,w. $$ From this expression, giving $ \alpha$ and $\beta $ we obtain $\Omega$ as function of $w$ and $\Phi$.}
of $\Phi.$  The corresponding covariant expression, defined as the
inverse $q_{\mu\nu} \, q^{\nu\lambda} = \delta^{\lambda}_{\mu}$, is
also a binomial expression\footnote{A metric of this form is also
relevant in another context. See for instance \cite{maqueijo}.}:
\begin{equation}
q_{\mu\nu} = \frac{1}{\alpha} \, \eta_{\mu\nu} -
\frac{\beta}{\alpha \, (\alpha + \beta) \, w} \, \partial_{\mu} \Phi
\, \partial_{\nu} \Phi.
\label{9junho11}
\end{equation}

Now we ask whether it is possible to find $\alpha $ and $ \beta,$ in such a way
that the dynamics of the field\ (\ref{23julho3}) takes the form
\begin{equation}
\Box \, \Phi = 0,
\label{23julho5}
\end{equation}
where $\Box$ is the Laplace-Beltrami operator relative to the metric $q_{\mu\nu}$, that is
$$ \Box \, \Phi \equiv  \frac{1}{\sqrt{- q}}
\partial_{\mu} ( \sqrt{- q} \,q^{\mu\nu} \,\partial_{\nu} \Phi).$$
To answer this question let us evaluate the determinant of the metric  $ q
= det \, q_{\mu\nu}.$ A direct calculation yields
\begin{equation}
\sqrt{- \, q} = \frac{\sqrt{- \,{\eta}}}{\alpha \, \sqrt{\alpha \,(\alpha + \beta)}}.
\label{lolo}
\end{equation}
Using equation (\ref{9junho1}), it follows that
\begin{equation} q^{\mu\nu} \, \partial_{\nu} \Phi   = \left( \alpha + \beta \right)  \,
\eta^{\mu\nu}\partial_{\nu}\Phi.
\label{ttt}
\end{equation}
The final result is summarized in the following:
\begin{theorem}
Given the Lagrangian $L = V(\Phi) w$ with an arbitrary potential $V(\Phi)$,
the field theory satisfying Eq. (\ref{23julho3}) in Minkowski spacetime is
equivalent to a massless Klein-Gordon field $\Box \Phi = 0$ in the metric
$q^{\mu\nu}$ provided that the functions $\alpha(\Phi)$ and $ \beta(\Phi) $ satisfy the condition
\begin{equation}
\alpha + \beta = \alpha^{3} \, V.
\label{9julho5}
\end{equation}
\end{theorem}

Remarkable, this equivalence is valid for any dynamics described in
the Minkowski background by the Lagrangian $L$. This fact can be
extended to other kinds of nonlinear Lagrangian (see details in
\cite{novelloetal2}).

\section{The general metric prescription: towards the geometrization of the scalar gravity}

We have shown how a nonlinear theory
based on the Lagrangian $L$ selects a class of metric tensors associated
with the dynamics of the scalar field. Is such property just a simple mathematical curiosity or is
the previous lemma pointing towards a more ambitious program?
Could the associated metric $q^{\mu\nu}$ play a more fundamental role?
The following
assumption takes a step in the direction of constructing
a theory for the gravitational interaction based on the
scalar field $\Phi$.

\begin{hypothesis}
The gravitational interaction is mediated by the scalar field $\Phi$.
All forms of matter and energy
interact with $\Phi$ only through the metric $q_{\mu\nu}$ and its derivatives in a covariant
way.
\end{hypothesis}

In the rest of the paper we explore this hypothesis to see whether it can stand
not only from the formal side but also in comparison with the  observations;
we will in particular examine the situation concerning the classical tests
of general relativity, cosmology and the gravitational radiation.

It is worthwhile to point out that the scalar field is not the
(special) relativistic generalization of the Newtonian potential.
Indeed, following the scheme of general relativity \cite{einstein}
and assuming that the test particles follow geodesics relative to
the geometry $q_{\mu\nu}$, we have that
\begin{equation}
\frac{d^{2} x^{i}}{dt^{2}} = - \, \Gamma^{i}_{00} = - \,
\partial^{i} \, \Phi_{N},
\label{22julho1}
\end{equation}
where we are assuming static weak field configuration and low
velocities for test particles.

From Eq. (\ref{9junho11}), we have
$$  \Gamma^{i}_{00}\approx - \, \frac{1}{2} \partial^{i} \, \ln
\alpha. $$ It follows that the Newtonian potential $ \Phi_{N} $ is
approximately given by
$$ \Phi_{N} \approx -\, \frac{1}{2} \, \ln \alpha,$$
which yields the relation between the $q-$metric and Newtonian potential $\Phi_N$ as
$$ q_{00} = \frac{1}{\alpha} \approx 1 + 2 \,\Phi_{N}.$$

Using equation (\ref{23julho5}) one obtains the right (vacuum)
Newtonian limit:
$$\nabla^2\Phi_N=0.$$
This was the starting point of Einstein\rq s path in building his
tensorial theory of gravity. The geometric scalar gravity follows
another path that we describe next. From now on we will explore the
consequences of extrapolating from the above approximation the
general expression

\begin{equation}
\label{alpha_phi}
\alpha = e^{- 2 \,\Phi}.
\end{equation}

The next task is to determine the functional dependence of $\beta$
on $\Phi$  or either the form of the potential $ V(\Phi)$ once
\begin{equation}
 \beta = \alpha \, ( \alpha^{2} \, V -
1). \label{18novembro} \end{equation}

Before doing this, a few comments on previous versions of scalar
theories are in order.

\section{The difficulties of previous scalar theories of
gravity}

One of the main drawbacks of the ancient proposals for scalar
gravity originates from the relation between the field $\Phi$ and Newtonian
potential $ \Phi_{N}$, while trying
to generalize Poisson\rq s equation \cite{giulini}.
Nordstr\"om was the first who tried a special relativistic generalization of
Newton's gravity \cite{nord}. He made the simplest assumption by imposing

\begin{equation}
\Box \, \Phi = -4\pi G\rho,  \label{24julho5}
\end{equation}
where $\rho$ is the matter density. Unfortunately the inertial mass
of the test particles was no longer constant and the theory could
not be derived from a variational principle. Afterwards, Einstein
himself \cite{coll-eins} stated in a very clear way the possibly
right generalization by writing the equation
\begin{equation}
\Box \, \Phi = -\kappa \, T, \label{24julho56}
\end{equation}
where $ T $ is the trace of the energy-momentum tensor. In this theory, test bodies move according with the equation
\begin{equation}
\frac{d }{d\tau} (m(x^{\alpha}(\tau))v^{\mu}(\tau)) =  m(x^{\alpha}(\tau))\eta^{\mu\nu}\partial_{\nu} \, \Phi(x^{\alpha}(\tau)),
\label{24julho7}
\end{equation}
where $\tau$ is the proper time. Note that there is a space-time dependent mass $m$ provided by

$$m=m_0e^{(\Phi-\Phi_0)},$$
where $m_0$ and $\Phi_0$ are constant. Einstein\rq s improved theory
has an action principle and the equivalence principle holds. However
the electromagnetic field does not couple to gravity. In these
lines, the best proposal was Einstein and Fokker's reformulation of
Nordstr\"om theory, in which they set

$$R=24\pi G\,T.$$
where $R$ is the curvature scalar. Yet, the electromagnetic field remains uncoupled to the gravitational field. Here is a summary of the main drawbacks of the above proposals:

\begin{itemize}
\item{\textit{Existence of a preferred-frame:} All forms of special relativistic theories accept this hypothesis from the very beginning.}
\item{\textit{The source of scalar gravity is the trace of the energy-momentum tensor:} this originates the main handicap of all previous scalar gravity: gravity does not couple to the electromagnetic field.}
\item{\textit{The scalar gravity is conformally flat:} Minkowski background is observable.}
\end{itemize}

Even after the advent of General Relativity and its successes in
accounting for observations, some alternative theories involving
scalar fields in different scenarios have been suggested up to now,
aiming to be competitive in explaining the observational tests and
sometimes bringing in new physics (see \cite{gibbons,giulini,finn}).

Our scalar theory of gravity overcomes the problems we mentioned above: the structure of the metric tensor\ (\ref{9junho1}) is covariant from the beginning, the field theory formulation makes the background metric unobservable and the action principle enables a full energy-momentum tensor coupling. It is also worth to emphasize the hypothesis that all bodies move along geodesics relative to the metric
$$q^{\mu\nu} = \alpha \, \eta^{\mu\nu} + \frac{\beta}{w} \, \partial^{\mu}\Phi \,
\partial^{\nu} \Phi.$$

As a realization of this procedure, in the next section we shall see how the electromagnetic fields couple to the gravitational field only through the metric $q_{\mu\nu}.$

\section{Electromagnetic field}

The equivalence principle states that all kind of matter and energy
interacts with the gravitational field trough minimal coupling
with the metric $ q_{\mu\nu}$. The old-fashioned scalar
gravity models assume that the scalar field only generates
a conformally flat metric and that it couples to the trace of the energy-momentum
tensor. These hypotheses are incorrect. In our geometric scalar
gravity (GSG) the electromagnetic field interacts with the scalar
field through the metric $q_{\mu\nu}$ as all other kind of matter and energy.

Accordingly the electromagnetic part of the Lagrangian is given by
\begin{equation}
L = F_{\alpha\mu} \, F_{\beta\nu}
\,q^{\mu\nu} \, q^{\alpha\beta}.
\label{7julho1}
\end{equation}
The corresponding field equation obtained by the variational
principle
$$ \delta \, \int \sqrt{- q}\,d^4x\, L  = 0,$$
is given by
\begin{equation}
F^{\mu\nu}{}_{; \, \nu} = 0, \label{17agosto1}
\end{equation}
where
$F^{\mu\nu} \equiv F_{\alpha\beta} \, q^{\alpha\mu} \,
q^{\beta\nu}$. The semicolon here represents
the covariant derivative w.r.t. the metric $q_{\alpha\beta}$.
The Hadamard conditions on discontinuities (see Appendix B) provide the dispersion relation
$$ k_{\mu} \, k_{\nu} \, q^{\mu\nu} =0,$$
where $k_{\mu}  \equiv \partial_{\mu} \Sigma$ is the gradient of the function $\Sigma$ that defines the
surface of discontinuity. Thus electromagnetic waves propagate along null geodesics of the geometry $ q_{\mu\nu}.$

We will now complete the theory by specifying parameters $ \alpha $ and $\beta$ as functions of $ \Phi.$
Following the strategy we have choose we have to look into observational conditions.
This is made first of all by looking into the gravitational field of spherical symmetric objects.

\section{The dynamics of scalar gravity}

The usual starting point to build a theory consists in writing an
action being guided by certain principles, such as general
covariance, second order differential equations, the right Newtonian
limit, and so on. A physical theory should however be falsifiable by
its observational consequences. Our geometric scalar gravity
contains two major ingredients: the interaction of the gravitational
field $ \Phi$ with matter and all kind of energy through the metric
tensor  $ q_{\mu\nu}$ and the dynamics of $\Phi$  related to the
Lagrangian $$L=V(\Phi)\, w$$  written in the auxiliary
(non-observable) Minkowski background. To select among all possible
Lagrangians of the above form we look for indications from the
various circumstances in which reliable experiments have been
performed. In this vein, we initiate the discussion by analyzing the
consequences of GSG for the solar system. This is of course not
enough, and to fully specify the dynamics we will need to look for
further properties. The analysis of planetary orbits allows however
to put constraints on the theory as we will now explain.

\subsection*{The static and spherically symmetric solution}

Any theory of gravity must account for planetary orbits. In general
relativity this motion is described by geodesics of the
Schwarzschild geometry. In the GSG particles follow geodesics in the
$q_{\mu\nu}$ metric.

Let us start by rewriting the (unobservable) auxiliary Minkowski
background metric in spherical coordinates
\begin{equation}
ds^{2}_{M} = dt^{2} - dR^{2}- R^{2} \, d\Omega^{2}.
\label{31julho01}
\end{equation}
Changing the radial coordinate to
$R = \sqrt{\alpha}\,r$,
where $\alpha=\alpha(r)$ we get
\begin{equation}
ds^{2}_{M} = dt^{2} - \alpha \left(\fracc{1}{2\alpha}\fracc{d\alpha}{dr}\,r+1\right)^2dr^{2}- \alpha r^{2} \, d\Omega^{2}.
\label{mink_mud}
\end{equation}
Since we are looking for static spherically symmetric solution we
assume that the field depends only on the radial variable $\Phi
=\Phi(r)$. Then the gravitational metric (\ref{9junho11}) takes the
form
\begin{equation}
ds^{2} = \frac{1}{\alpha} \, dt^{2} - B \,dr^{2}- r^{2} \, d\Omega^{2},
\label{31julho1}
\end{equation}
where we have defined
$$ B \equiv \frac{\alpha}{\alpha+\beta}\left(\fracc{1}{2\alpha}\fracc{d\alpha}{dr}\,r+1\right)^2.$$

The field equation (\ref{23julho5}) then reduces to
\begin{equation}
\fracc{r^2\sqrt{\alpha +\beta}}{\alpha}\left(\fracc{1}{2\alpha}\fracc{d\alpha}{dr}\,r+1\right)^{-1} \, \frac{d\Phi}{dr} = \Phi_0,
\label{31julho31}
\end{equation}
where $\Phi_0$ is a constant. Noting that $ \alpha(\Phi) = e^{-2
\Phi}$ we proceed by successive approximations (see details in
 Appendix C) and get an ansatz for the form of the
potential $ V$ as follows:
\begin{equation}
V(\Phi) = \fracc{(\alpha-3)^2}{4 \, \alpha^3}
\label{alp_sch}.
\end{equation}

By substituting Eqs.\ (\ref{alpha_phi}) and\ (\ref{alp_sch}) into Eq. (\ref{31julho31})
gives
\begin{equation}
\label{dyn_subs_ans}
e^{2\Phi}\,\frac{d\Phi}{dr}=\fracc{\Phi_0}{r^2}.
\end{equation}
and therefore
\begin{equation}
\Phi=\fracc{1}{2}\ln\left(2c_1-2\fracc{\Phi_0}{r}\right),
\end{equation}
where $c_1$ is an integration constant. The asymptotic behavior
implies that $c_1=1/2$ and $\Phi_0 = MG/c^2$, where $M$ is the mass
of gravitational source, $G$ is Newton's constant and $c$ the speed
of light, i.e.
\begin{equation}
\Phi=\fracc{1}{2}\ln\left(1-\fracc{r_H}{r}\right),
\end{equation}
where $r_H\equiv2MG/c^2$. The scalar field $\Phi$ reduces to
Newtonian potential in the weak field limit. Using $ \alpha$ and $
\beta $ given by\ (\ref{alpha_phi}) and (\ref{18novembro}),
respectively, the line element can be written as
\begin{equation}
\label{line_el_scha}
ds^2 = \left(1-\frac{r_H}{r}\right)dt^2 - \left(1-\frac{r_H}{r}\right)^{-1}dr^2 - r^2d\Omega^2.
\end{equation}
This geometry has the same form as in general relativity and yields
the observed regime for solar tests. Thus, the present geometric
scalar gravity is a good description of planetary orbits and also
for light rays trajectories that follow geodesics (time-like and
null-like, respectively) in the $q_{\mu\nu} $ geometry. If new
observations would require a modification of the metric in the
neighborhood of a massive body this should be made by adjusting the
form of the potential $ V(\Phi).$

\section{Action principle}

Now that we have found a possible potential for the scalar field we
are in position to write its dynamical equations. Let us start by
the action written in the auxiliary unobservable Minkowski
background. From variational principle
$$ \delta  \, S_{1}= \delta \, \int \sqrt{- \eta} \,d^4x\, L, $$
we get:
\begin{eqnarray}
\delta \, S_{1} &=& - \, \int \, \sqrt{- \eta} \, d^4x\, \left( V' \, w +
2 \, V \, \Box_{M} \Phi \right) \, \delta\Phi
\end{eqnarray}
where $$\Box_{M} \Phi\equiv\frac{1}{\sqrt{-\eta}}\partial_{\mu} \,
\left(\sqrt{-\eta}  \,\eta^{\mu\nu}\, \partial_{\nu} \Phi\right)$$
is the d\rq Alembert operator in flat space (in general
coordinates). Using the bridge relations to pass to curved
space-time we get
\begin{equation}
\delta \, S_{1} = - \, 2 \, \int   \sqrt{- q} \, d^4x \, \sqrt{V} \,\Box
\Phi  \, \delta \Phi.
\end{equation}
In presence of matter we add a corresponding term $L_{m}$ to the
total action:
\begin{equation}
S_{m} = \int \, \sqrt{- q} \, d^4x\, L_{m}.
\label{24julho15}
\end{equation}
The first variation of this term as usual yields
\begin{equation}
\delta S_{m} = - \, \frac{1}{2} \, \int \, \sqrt{- q} \, d^4x\, T^{\mu\nu}
\, \delta \, q_{\mu\nu}, \label{24julho17}
\end{equation}
where we have defined the energy-momentum tensor in the standard way
$$ T_{\mu\nu} \equiv  \, \frac{2}{\sqrt{- q}} \,
\frac{\delta( \sqrt{- q} \, L_{m})}{\delta q^{\mu\nu}}.$$ General
covariance leads to conservation of the energy-momentum tensor $
T^{\mu\nu}{}_{;\nu}=0.$ The equation of motion is obtained by the
action principle
$$\delta S_{1} + \delta S_{m}= 0.$$
Up to this point we are following the paths of general relativity.
Here however, in the GSG theory, the metric $ q_{\mu\nu} $ is not
the fundamental quantity. We have to write the variation $ \delta
q_{\mu\nu}$ as function of $ \delta \Phi:$
\begin{equation}
\delta \,
q_{\mu\nu} = \delta \left(\frac{1}{\alpha} \, \eta_{\mu\nu} - \frac{\beta}{\alpha Z w} \, \partial_{\mu} \, \Phi \, \partial_{\nu} \,\Phi\right).
\end{equation}
While we are using the unobservable background Minkowski metric for
simplicity, at the end all expressions should be written in terms of
the gravitational metric $q_{\mu\nu}$. After some calculation we get

$$ \fracc{\delta S_{m}}{\delta\Phi} = - \frac{1}{2} \int \sqrt{- q} \,d^4x \left( \frac{\alpha'}{\alpha} (E-T) - \frac{Z'}{Z}E + 2 \nabla_{\lambda} C^{\lambda}\right), $$
where $Z\equiv\alpha+\beta = \alpha^{3} \, V$. We have also denoted
$$ T\equiv T^{\mu\nu} \, q_{\mu\nu}, \hspace{.3cm} E \equiv \frac{T^{\mu\nu} \, \partial_{\mu}\Phi \, \partial_{\nu}\Phi}{\Omega}, \hspace{.3cm} X' \equiv \frac{dX}{d\Phi},$$
and
$$ C^{\lambda}\equiv\frac{\beta}{\alpha \, \Omega} \, \left( T^{\lambda\mu} - E \, q^{\lambda\mu} \right) \, \partial_{\mu}\Phi .$$
Finally, the equation of motion for the gravitational field $ \Phi$
takes the form:
\begin{equation}
\sqrt{V} \, \Box\Phi= \kappa \, \chi, \label{12out1}
\end{equation}
where
$$
\chi = \fracc{1}{2} \, \left( \frac{\alpha'}{2\alpha} \, (T-E) +
\frac{Z'}{2Z}E - \nabla_{\lambda} \, C^{\lambda}\right).
$$
Substituting the value $ \alpha = e^{- 2 \Phi}$ and using the equation
(\ref{alp_sch}) for the potential $V$ we rewrite this expression
under the form

$$
\chi = \fracc{1}{2} \, \left( \frac{3 \, e^{2 \Phi} + 1}{3 \,
e^{2\Phi}- 1} \, E - T - \nabla_{\lambda} \, C^{\lambda}\right).
$$

This equation describes the dynamics of our GSG in presence
of matter, under the assumptions\ (\ref{alpha_phi}) and\ (\ref{alp_sch}). The quantity $\chi$
involves a non-trivial coupling between the gradient of the scalar
field  $\nabla_{\mu} \Phi$ and the complete energy-momentum tensor
of the matter field $T_{\mu\nu}$  and not uniquely its trace. This
property allows the electromagnetic field to interact with the
gravitational field. The Newtonian limit gives the identification
$$\kappa\equiv\fracc{8\pi G}{c^4}.$$

\subsection*{Natural decomposition}

The form of the metric, containing the derivative of $ \Phi $
suggests a simplification in the description of the matter terms
which is useful for exploring the cosmological consequences of GSG.
Suppose that $\partial_{\mu} \Phi$ is time-like, that is $ \Omega
> 0$. We then define the normalized vector

\begin{equation}
I_{\mu} = \frac{\partial_\mu \Phi}{\sqrt{\Omega}}.
\end{equation}

This vector can be used to decompose the energy-momentum tensor of a
perfect fluid in the "co-moving" representation by setting

\begin{equation}
T^{\mu\nu} = ( \varrho + p) \, I^{\mu}  \, I^{\nu}  - p \, q^{\mu\nu},
\end{equation}
it then follows
$$  T^{\mu\nu} \, \partial_{\mu}\Phi = \sqrt{\Omega} \, \varrho \, I^{\mu}, $$
and
$$  T^{\mu\nu} \, \partial_{\mu}\Phi \, \partial_{\nu}\Phi = \Omega \, \varrho.  $$

Thus, in this frame the quantities $ E $ and $ T $ reduces to
\begin{equation}
E = \varrho, \hspace{1cm} T = \varrho - 3 p.
\end{equation}
Using these results it follows that $ C^\mu = 0.$ In the natural
frame associated to the gradient of the gravitational field $ \Phi $
the equation of motion for the scalar gravity, reduces to the form
\begin{equation}
\sqrt{V} \, \Box \, \Phi =  - \frac{\kappa}{2} \,  \left( \frac{2 \,
\alpha}{\alpha-3} \, \varrho - 3 p \right).
\end{equation}
This is the form of the dynamics of $ \Phi$ when the source is a
perfect fluid. In the next section we provide a simple example of GSG in the analysis of the global properties of the universe.

\section{Cosmology}
In this section we start discussing some cosmological aspects of
Geometric Scalar Gravity.

We work in a Gaussian coordinate system that provides a 3 + 1
decomposition of the spacetime manifold. Because of homogeneity and
isotropy, the gravitational field depends only on the global time T:
$$ \Phi= \Phi(T).$$ We set for the auxiliary (non-observable)
Minkowski background the expression
\begin{equation}
ds^{2}_{M} = dT^{2} - dx^{2} - dy^{2} - dz^{2}
\end{equation}
Then the gravitational metric is given by
\begin{equation}
q^{00} = Z, \,\, q^{11} = q^{22} = q^{33} = - \,\alpha,
\end{equation}
where $ \alpha = \alpha(T)$ and $ Z = Z(T).$Then the gravitational metric takes the form
$$ ds^{2} = \fracc{1}{Z}\, dT^{2} - \fracc{1}{\alpha} \, (dx^{2} + dy^{2} +
dz^{2}).$$ After making some redefinitions  $\alpha\equiv1/a^2$ and
$dt=dT/\sqrt{Z}$, the line element reduces to

\begin{equation}
ds^{2} = dt^{2} - a(t)^{2} \, (dx^{2} + dy^{2} + dz^{2}).
\label{30juho1}
\end{equation}
We use the natural frame defined in the previous section to write
the source of the gravitational field as a perfect fluid with
equation of state $p=\lambda\rho$. Using Eq.\ (\ref{alpha_phi}) it follows that $a=e^{\Phi}.$ Therefore, the dynamical equation for the gravitational field $ \Phi$ becomes an equation of motion in terms of the scale factor $a(t)$:
\begin{equation}
\sqrt{V} \, \left(\fracc{\ddot a}{a}+2\fracc{\dot a^2}{a^2} \right)
=  - \, \fracc{\kappa}{2} \, \left( \frac{2}{ 1 - 3 \, a^{2}}
\,\varrho - 3 \, p \right),
\label{dyn_sc_fac}
\end{equation}
where $\dot X\equiv dX/dt$. On the other hand, the equation of
conservation of the energy-momentum tensor yields

\begin{equation}
\varrho=\varrho_0\,a^{-3(1+\lambda)}.
\label{cont}
\end{equation}
Using this formula in Eq.\ (\ref{dyn_sc_fac}), yields
\begin{equation}
\fracc{2}{\kappa} \, \sqrt{V} \, \left( \fracc{\ddot
a}{a}+2\fracc{\dot a^2}{a^2} \right) = \, - \, \fracc{2 -3 \, \lambda +
9 \, \lambda \, a^{2}}{ 1 - 3 \, a^{2}} \, \varrho_0 \,
a^{-3(1+\lambda)}.
\label{16novembro}
\end{equation}

We should compare this behavior of the scalar factor of the
cosmological metric with the corresponding equations obtained in
general relativity. In GR there are two equations. A dynamical one
\begin{equation}
\frac{\ddot A}{A}=-\fracc{k}{6}(1+3\lambda)\varrho,\label{eq_fried1}
\end{equation}
and a constraint

$$ \left(\fracc{\dot
A}{A}\right)^{2}=\fracc{k}{3}\varrho $$ where $A$ plays the role of $a(t)$
in these equations.

Equation\ (\ref{eq_fried1}) corresponds to the true dynamical one
and the other one is nothing but restricts the initial condition.
Therefore, the integration constants of Eq.\ (\ref{eq_fried1}) are
not arbitrary. There is no analogue constraint equation in GSG. We
postpone the further analysis of the cosmological description of the
scalar gravity for a future work.

 \section{Conclusion and Outlook}

 Before proposing to jump from one to ten quantities for
 describing the gravitational field, Einstein and other contemporary researchers
 tried the natural modification from one function as used in Newtonian gravity
 to one scalar (special relativistic) field. However, all these
 attempts failed.

 Einstein then came with his beautiful idea that gravity is a metrical
 phenomenon and the jump from one to ten functions became natural.
 Besides, the success of general relativity made this excess of
 variables not a weak point but, on the contrary, it appeared as
 necessary.

 In the present paper, we have clarified many facets of the problem and showed how it
 is possible to construct a metrical theory of gravity (Einstein's
  insight) avoiding the necessity to increase the number of
 fundamental variables in the Geometrical Scalar Gravity.

 Our proposal deals with gravitational
 processes through a scalar field in a natural extension of Newton\rq
 s theory. The traditional drawbacks of scalar gravity are overcome.
 The auxiliary Minkowski geometry $\eta_{\mu\nu}$ is not
 observable. This procedure is similar to the background field
 formulation of General Relativity proposed by Gupta, Feynman, Deser,
 Grischuk and others. Matter interacts with the field $ \Phi$ only
 through the gravitational metric

$$q^{\mu\nu} = \alpha \, \eta^{\mu\nu} + \frac{\beta}{w} \, \partial^{\mu} \Phi \, \partial^{\nu} \Phi, $$
where $$ \alpha = e^{-2\Phi},$$
$$ \beta = \alpha \, (\alpha^{2} \, V - 1).$$

The fundamental Lemma relates a class of dynamical systems described
by the Lagrangian $L = V(\Phi) \, w $ to a massless Klein-Gordon
self-interacting field in a curved spacetime.

 The self-interaction of the scalar field is described by the potential
 $V(\Phi)$ which acts multiplicatively on the kinetic term. Different
 forms of $V(\Phi)$  yields different gravitational theories. In the
 present paper we analyzed the case in which $$V = \frac{(\alpha -
 3)^{2}}{4 \, \alpha^{3}}. $$ Such particular choice allows solutions
 in agreement with actual observational knowledge of gravity (see
 appendix C). The crucial issue of  gravitational radiation and in particular the emission
 of gravitational energy from the compact sources will be analyzed in a future paper.

 In  appendix A, we exhibit a preliminary analysis based on Ref.
 \cite{shapiro}, in which we show that the rate of energy emitted by a monopole
 radiation in our GSG has the same order of magnitude of the
 quadrupole radiation in general relativity. This and the cosmological consequences of Geometric Scalar
 Gravity deserve a more specific analysis that will be described
 elsewhere.

\section{APPENDIX A: Gravitational radiation}

As mentioned before, a theory of gravity should be prepared to account for the observed
behavior of binary pulsars as a consequence of the emission of
gravitational energy (for instance, see \cite{taylor}). General
relativity makes use of the quadrupole radiation formula with
success. Let us analyze this phenomenon from the point of view of
GSG.

We follow the main lines as presented in \cite{shapiro} and
calculate the equation of motion of the scalar field $\Phi$ in the
Minkowski background under the weak field regime. We provide then an
estimative of the order of magnitude of the gravitational waves in
the radiation zone using GSG theory.

We bound the discussion to the simplest case of a perfect fluid in
the natural frame. This hypothesis implies that $C^{\mu}=0$. Under
this consideration, the equation of motion is

\begin{equation}
\label{gw_eq} \sqrt{V} \, \Box \, \Phi = - \frac{\kappa}{2} \,  \left(
\frac{2 \, \alpha}{\alpha-3} \, \varrho - 3 p \right).
\end{equation}
Use of the Lemma allows to rewrite Eq.\ (\ref{gw_eq}) in the
Minkowski background as follows

\begin{equation}
\Box_M\Phi + \frac{1}{2} \, \frac{V'}{V} \, w   = \fracc{k}{2} \,
\left(\frac{1}{\alpha \, \sqrt{V}}\right)^{3} \,
\left(3p-\fracc{2\alpha}{\alpha-3}\varrho\right). \label{gw_back}
\end{equation}

Let us assume that the source of the gravitational field is dust
($p=0$) and rewrite the energy density in terms of background
quantities $\varrho=\alpha\varrho_b $ where $\varrho_b$ is the
energy density in the Minkowski space-time. Inserting the
expressions of $V$ and $\alpha$  in terms of $\Phi$ the equation of
motion for the gravitational field is

\begin{equation}
\Box_M\Phi=\fracc{k}{2} \left[ \fracc{2}{k}
\fracc{\alpha-9}{\alpha-3}\, w - 2 \, \left(\frac{1}{\alpha \,
\sqrt{V}}\right)^{3} \,
\fracc{\alpha^2}{\alpha-3}\,\varrho_b\right]. \label{gw_back3}
\end{equation}
This equation of motion has an extra term proportional to $w$ which
involves first derivatives of the scalar field. To proceed and
calculate the total rate of energy emission for a spherically
symmetric energy distribution, we have to take into account
conservation laws. Matter conservation implies

\begin{equation}
M_0=\int d^3x\,\gamma\,\varrho_b=const,
\label{rest_mass}
\end{equation}
where $\gamma\equiv[1-(v/c)^2]^{-1}$ is the Lorentz factor. This formula is responsible to the non-radiative term of the scalar field as we shall see.

The energy-momentum tensor of the scalar field in the Minkowski background is given by

\begin{equation}
T^{\mu\nu}_{\Phi}=\fracc{2V}{k}\left(\partial^{\mu} \Phi \,
\partial^{\nu} \Phi-\fracc{\omega}{2}\eta^{\mu\nu}\right).
\label{t_munu_phi_mink}
\end{equation}
This expression will be useful to calculate the rate of radiative
emission. By applying the Green method to solve Eq.\
(\ref{gw_back3}), the scalar field $\Phi$ takes the form

\begin{equation}
\Phi(t,{\bf x})=-G \,\int{d^3x'}\fracc{\left[\tilde\varrho\right]_{ret}}{|{\bf x}-{\bf x'}|},
\label{pot_ret}
\end{equation}
where $[X]_{ret}$ means the evaluation of $X$ in the retarded time
$t-|{\bf x}-{\bf x'}|$ and
$$\tilde\varrho \equiv\left[ - \fracc{2}{k} \fracc{\alpha-9}{\alpha-3}\, w + 2 \,\left(\frac{1}{\alpha \, \sqrt{V}}\right)^{3}
\, \fracc{\alpha^2}{\alpha-3}\,\varrho_b\right].$$ We define the
rest density by

\begin{equation}
\varrho_0\equiv\gamma\varrho_b.
\label{rest_rho}
\end{equation}
Then, in the wave zone regime ($r=|{\bf x}|$ and $r\approx r'\cos\theta'$), we get

\begin{eqnarray}
\label{pot_ret_wz} \Phi(t,{\bf x})&=& - \fracc{G}{r} \left(
\int{d^3x'\left[2 \, \left(\frac{1}{\alpha \, \sqrt{V}}\right)^{3}
\, \fracc{\alpha^2}{\alpha-3}\fracc{\varrho_0}{\gamma}\right]_{ret}}+\right.\nonumber\\[2ex]
&&\left.- \fracc{2}{k}\int{d^3x'\left[\fracc{\alpha-9}{\alpha-3}\left(\dot\Phi^2-(\nabla\Phi)^{2}\right)\right]_{ret}}\right)+...\nonumber\\[2ex]
\end{eqnarray}
Due to the partial derivatives of the scalar field, the second
term on the right hand side of this equation is of second order in $\Phi$ and can be dropped out.
Expanding the remaining term in Taylor series, we obtain

\begin{eqnarray}
\varrho_0(t',{\bf x'})&=&\varrho_0(t-r,{\bf x'})+r'\cos\theta'\dot\varrho_0 +\nonumber\\[2ex]
&&+\fracc{1}{2} r'^2\cos^2\theta'\ddot\varrho_0+...
\label{exp_rho}
\end{eqnarray}
and

\begin{equation}
\fracc{2}{\gamma} \left(\frac{1}{\alpha \, \sqrt{V}}\right)^{3}
 \, \fracc{\alpha^2}{\alpha-3}=-\left[1-11\Phi-v^2\right]_{t-r}+...
\label{exp_gamma}
\end{equation}

Therefore, the leading-order radiative term in the spherically symmetric case is

\begin{equation}
\Phi(t,r)= - \fracc{4\pi G}{r} \left\{ \int{dr'r'^2\left[ \varrho_0 \left(11\Phi+v^2\right)\right]_{ret}} + {\cal O}\,(2) \right\},
\label{pot_ret_ss}
\end{equation}

Now we use Eq.\ (\ref{pot_ret_ss}) and the virial relation $v^2\sim GM/R$ to estimate $\Phi$ for a
source of mass $M$, radius $R$ and velocity $v$, in the weak-field regime:

\begin{equation}
\Phi\sim \fracc{GM}{rc}\left(\fracc{v}{c}\right)^2.
\label{est_phi}
\end{equation}

The definition of the total rate of energy emission in the wave zone is given by

\begin{equation}
\fracc{d{\cal E}}{dt}=-4\pi r^2T^{0r}_{\Phi}\approx-\fracc{1}{G} (r\dot\Phi)^2,
\label{rate}
\end{equation}
where we use $d\Phi/dr=-d\Phi/dt$ and Eq.\ (\ref{t_munu_phi_mink}) to express the radiative energy flux
\begin{equation}
T^{0r}_{\Phi}=\fracc{1}{4\pi G}\dot\Phi^2.
\nonumber
\end{equation}

Finally, using the correct units, we get

\begin{equation}
\fracc{d{\cal E}}{dt}\sim\fracc{c}{G} \Phi^2\sim\fracc{c}{G} \left(\fracc{GMv^2}{rc^3}\right)^2\sim\fracc{G}{c^5}\left(\fracc{Mv^2}{r}\right)^2.
\label{rate_est}
\end{equation}

It is clear that a very careful analysis on this topic should be
done. The aim of this section is just to show that the non-linear
scalar theory of gravity presented here provides a monopole
radiation for a spherically symmetric mass distribution in the GSG
of the same order of magnitude of the quadrupole gravitational
radiation $G/c^5$ as in general relativity. A more precise
evaluation of the binary pulsar timing is a non-trivial matter which
we intend to do in the near future.

\section*{APPENDIX B: Hadamard Discontinuity}
Let us analyze the discontinuities of the electromagnetic field in
our scalar theory of gravity. We shall show that the electromagnetic
waves will follow a geodesic motion in $q_{\mu\nu}$.

We proceed according to the standard Hadamard method (see details in
\cite{novelloetal3} and references therein) and obtain the dispersion
relation of the waves. Let $\Sigma$ be a surface of discontinuity of
the field $A_{\mu}$. The discontinuity of an arbitrary function $f$
is given by:
\begin{equation}
\label{DiscDef}
{\left[ f(x) \right]}_\Sigma = \lim_{\epsilon \to 0^+} \bigl( f(x +
\epsilon) - f(x - \epsilon)\bigr).
\end{equation}
The field $A_{\mu}$ and its first derivative $\partial_\nu A_{\mu}$ are
continuous across $\Sigma$, while the second derivatives presents a
discontinuity:
\begin{eqnarray}\label{CondHadamard}
\left[ A_{\mu} \right]_\Sigma &=& 0,\\
\left[ \partial_\nu A_{\mu} \right]_\Sigma &=& 0,\\
\left[ \partial_\alpha\partial_\beta A_{\mu} \right]_\Sigma &=& k_\alpha k_\beta
\xi_{\mu}(x),
\end{eqnarray}
where $ k_\mu \equiv \partial_\mu \Sigma$ is the propagation vector and
$\xi_{\mu}(x)$ the amplitude of the discontinuity. Using these
discontinuity properties in the equation of motion
$$q^{\alpha\nu}A_{[\mu,\nu];\alpha}=0,$$ it follows that:
\begin{equation*}
k_\alpha k_\beta q^{\alpha\beta}=0.
\end{equation*}
This means that the discontinuities of the electromagnetic field
propagate as null geodesics in the metric $q_{\mu\nu}$. It happens
due to the relation between the EM field and the metric, which is a
minimal coupling from the general relativity point of view.
Therefore, any other kind of matter will see only the metric
$q_{\mu\nu}$.

\section*{APPENDIX C: The form of the potential V}
From the relation imposed by the Lemma
$$ V = \frac{Z}{\alpha^{3}}, $$ it follows that the observation of $
Z$ specifies the potential. From equation (\ref{31julho1}) it
follows that \begin{equation}
-q_{11} \equiv B = \frac{\alpha \, (\alpha - 3)^{2}}{4Z}.
\label{A1}
\end{equation}

Observations have been made up to second order on $ x \equiv
r_{H}/r.$ This means that we can be confident that up to this order
the form of the $ q_{11}$ coefficient of the gravitational metric is
given by

\begin{equation}
 B \approx 1 + x + x^{2}.
 \label{A2} \end{equation}
This implies that one can be sure on
the expansion of $ Z $ up to second order on $ \alpha.$ Any further
dependence on $ \alpha$ should be impossible to be univocally fixed
in this order of expansion. Indeed, if we set
$$ Z = M \, \alpha^{2} + N \, \alpha + P, $$
substituting in this expression and expanding all quantities up to
second order $ O(x^{2})$ and noting that $$ q_{00} =
\frac{1}{\alpha} = 1 - x$$ yields $$\alpha \approx 1 + x + x^{2},$$
it follows
$$ M = \frac{1}{4}; \hspace{.3cm} \,N = - \, \frac{3}{2}; \hspace{.3cm} P =
\frac{9}{4 }.$$
In other words, this gives
$$ Z = \frac{( \alpha - 3 )^{2}}{4}$$
and consequently
$$ V = \frac{( \alpha - 3)^{2}}{4 \, \alpha^{3}}.$$
This corresponds to the expression (\ref{alp_sch}).\\

\section*{APPENDIX D: Geometrical version of Einstein-Nordstr\"om theory}
Let us consider now the particular case in which the potential takes
the form
$$ V = \frac{1}{\alpha^{2}}, $$
and set
$$ \alpha = e^{2\, \Phi}.$$

\vspace{0.5cm}

As a consequence, the coefficient $ \beta$ of the
metric $q_{\mu\nu}$ vanishes. The form of the metric reduces to a
conformally flat geometry, that is
$$ q_{\mu\nu} = e^{- \, 2 \, \Phi} \, \eta_{\mu\nu}.$$
The equation of motion (\ref{12out1}) is the Einstein-Nordstr\"om
dynamics
$$ \Box\Phi= - \, \frac{\kappa}{2} \,e^{2\Phi} \, T$$
that is
$$ \Box\Phi= - \, \frac{\kappa}{2} \, T^{\mu\nu} \, \eta_{\mu\nu}$$
once  $ T = T^{\mu\nu} \,q_{\mu\nu} =  T^{\mu\nu} e^{-
\,2 \, \Phi} \, \eta_{\mu\nu}.$ A direct comparison with the limit
of weak field as presented in a previous section shows that this is
not a good proposal once it describes configurations which are not
compatible with the weak gravitational field. The theory constructed
with such bad choice of the potential $V$ is nothing but the
unsuccessful Einstein-Nordstr\"om scalar gravity.

\section*{Acknowledgements}
MN and EB would like to thank ICRANet (Pescara) and Universit\`a
degli Studi dell'Insubria (Como) for their hospitality where part of
this paper was written. This work was supported by {\em Conselho
Nacional de Desenvolvimento Cient\'{\i}fico e Tecnol\'ogico} (CNPq),
FINEP and {\em Funda\c{c}\~ao de Amparo \`a Pesquisa do Estado do
Rio de Janeiro} (FAPERJ) of Brazil.

\end{document}